\documentclass[11pt,twoside]{article}

\usepackage{asp2006}
\usepackage{epsf}
\usepackage{lscape}
\usepackage{graphicx}

\begin{document}
\title{Numerical Modeling of AGN Jets: Formation of Magnetically
Dominated Lobes and Stability Properties of Current-carrying Jets}   
\author{M. Nakamura, H. Li, S. Diehl, and S. Li}   
\affil{Theoretical Division, Los Alamos National
Laboratory, Los Alamos, NM 87545, U.S.A}    

\begin{abstract}
We argue  the behavior of  Poynting flux-dominated outflows from  AGN in
the  galactic  cluster   systems  by  performing  three-dimensional  MHD
simulations  within the  framework of  the "magnetic  tower"  model.  Of
particular  interests are the  structure of  MHD waves,  the cylindrical
radial force balance, the  (de)collimation, and the stability properties
of  magnetic  tower  jets.   Transition  between the  jet/lobe  and  the
formation  of wiggling  jet  by growing  current-driven instability  are
discussed.
\end{abstract}

\section{Introduction}
Magnetohydrodynamic (MHD)  mechanisms are  often invoked to  explain the
launching,  acceleration  and collimation  of  jets  from Young  Stellar
Objects,  X-ray binaries, Active  Galactic Nuclei  (AGNs), Microquasars,
and  Quasars \citep[see,  {\it  e.g.},][and references  therein]{DLM01}.
Strongly  magnetized jets,  particularly  those with  a strong  toroidal
field  encircling  the  collimated   flow,  are  often  referred  to  as
``current-carrying'' or ``Poynting flux-dominated'' (PFD) jets.  A large
current flowing parallel to the jet flow is responsible for generating a
strong, tightly wound  helical magnetic field.  The global  picture of a
current-carrying  jet   with  a   closed  current  system   linking  the
magnetosphere of the central engine  and the hot spots was introduced by
\citet[]{B78, B06} and applied to AGN double radio sources.  This closed
current system includes a pair of current circuits, each containing both
a forward electric current path  (the jet flow itself, with its toroidal
magnetic field,  toward the  lobe), and a  return electric  current path
(along some  path back to the  AGN core). 

Theory of magnetically driven outflows in the electromagnetic regime has
been  proposed  by  \citet[]{B76}  and  \citet[]{L76}  and  subsequently
applied  to  rotating  black  holes  \citep[]{BZ77}  and  to  magnetized
accretion  disks  \citep[]{BP82}.   An underlying  large-scale  poloidal
field for  producing the magnetically driven jets  is almost universally
assumed in  many theoretical/numerical models.  However,  the origin and
existence  of   such  a  galactic   magnetic  field  are   still  poorly
understood. In  contrast with the large-scale  field models, Lynden-Bell
\citep[]{L96} examined  the ``magnetic  tower''; expansion of  the local
force-free magnetic loops anchored to the star and the accretion disk by
using  the semi-analytic  approach.  Global  magnetostatic  solutions of
magnetic  towers with external  thermal pressure  were also  computed by
\citet[]{Li01}  using the Grad-Shafranov  equation in  axisymmetry. Full
three-dimensional MHD numerical simulations of magnetic towers have been
performed by \citet[]{K04b}.

Recent  X-ray  and  radio   observations  have  revealed  the  dynamical
interaction  between the  outbursts  driven by  AGN  and the  background
IGM/ICM, such  as X-ray  "cavities" with corresponding  radio "bubbles''
\citep[]{F00,  C03, K06,  W07, G07}.   The cavities  are believed  to be
filled  with very  low density  relativistic plasma,  inflated  from the
electromagnetic  jets  that  are  being  powered by  AGNs.   This  paper
describes  nonlinear  dynamics of  propagating  magnetic  tower jets  in
galaxy cluster scales  ($>$ tens of kpc) based  on three-dimensional MHD
simulations  to  argue the  jet/lobe  transition  \citep[]{N06} and  the
stability properties \citep[]{N07}.

\section{Numerical Setup}
We  solve the  nonlinear system  of time-dependent  ideal  MHD equations
numerically  in a  3-D Cartesian  coordinate system  $(x,\,y,\,z)$.  The
basic numerical  treatments are  introduced in \citet[]{L06,  N06, N07}.
We  assume an  initial  hydrostatic equilibrium  in the  gravitationally
stratified medium,  adopting an iso-thermal King  model \citep[]{K62} to
model  the magnetic  towers from  AGNs in  galaxy cluster  systems.  AGN
accretion disk can not be resolved in our computational domain, and thus
the magnetic flux and the mass  are steadily injected in a central small
volume during a certain time period.  Since the injected magnetic fields
are  not  force-free, they  will  evolve  as  a ``magnetic  tower''  and
interact with the  ambient medium. In the present  paper, we present two
different runs: one is called  the ``unperturbed case'' in the following
discussion,  which is  a run  without  any initial  perturbation to  the
background  initial profiles  \citep[]{N06}.   The other  is called  the
``perturbed case'', where a finite amplitude perturbation (a few percent
of  the  background sound  speed)  is given  to  the  velocities of  the
background gas \citep[]{N07}.

The total computational domain is  taken to be $|x|,\,|y|,\,|z| \leq 16$
corresponding to a  (160 kpc:$-80$ to $80$ kpc)$^3$  box.  The numerical
grids are $240^3$  in the unperturbed case and  $320^3$ in the perturbed
case.  Normalizing  factors are a  length $R_0 =  5$ kpc, a  sound speed
$C_{\rm s0} =  4.6 \times 10^7$ cm s$^{-1}$, a time  $t_{0} = 1.0 \times
10^7$ yr, a density $\rho_{0}  = 5.0 \times 10^{-27}$ g cm$^{-3}$.  The
corresponding unit pressure $p_0$ as  $ \rho_0 C_{\rm s0}^2 = 1.4 \times
10^{-11}$ dyn cm$^{-2}$, and the unit magnetic field $B_{0}$ as $\sqrt{4
\pi \rho_0 C_{\rm  s0}^2}=17.1$ $\mu$G.  The initial sound  speed in the
simulation   is  constant,   $C_{\rm   s}=\gamma^{1/2}  \approx   1.29$,
throughout the  computational domain, which  give a sound  crossing time
$\tau_{\rm s}  \approx 0.78$,  corresponding to a  typical time  scale $
R_{0} / C_{\rm  s0} \approx 10.0$ Myr.  Therefore,  $t=1$ is equivalent
to the unit time  scale $12.8$ Myr.  In the King model  we use here, we
adopt the cluster core radius $R_{\rm c}$ to be $4.0$ (i.e., 20 kpc) and
the slope $\kappa$ to be $1.0$ in the unperturbed case and $0.75$ in the
perturbed case.  Magnetic fluxes and mass are continuously injected into
a central  volume of the computational  domain for $t_{\rm  inj} = 3.1$,
after which  the injection is  turned off.  A magnetic  energy injection
rate is  $\sim 10^{43}$  ergs s$^{-1}$, a  mass injection rate  is $\sim
0.046 M_\odot$/yr, and an injection time is $\sim 40$ Myr.

\section{Results}
\begin{figure}[!h]
\begin{center}
\includegraphics[scale=0.8,            angle=0,            clip]{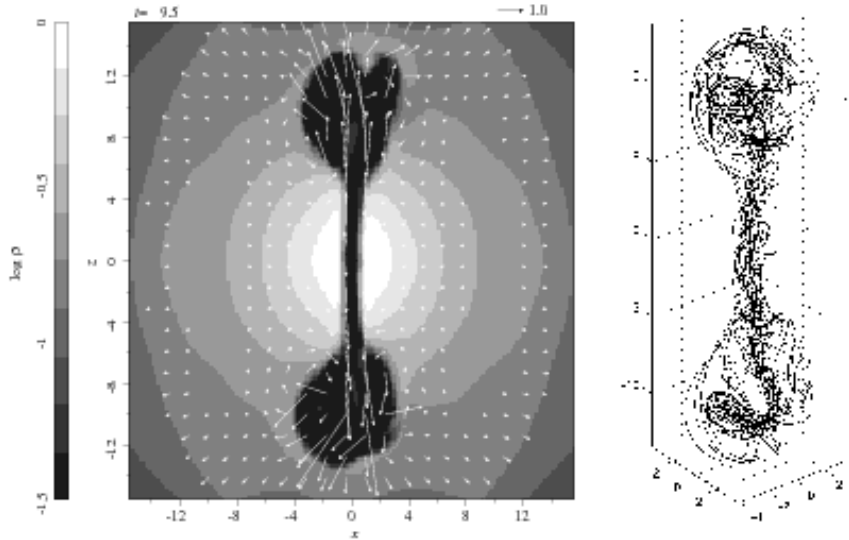}
\caption{\label{fig:f1} Distribution of density in the $x-z$ plane along
with the  poloidal velocity ({\em  left}) and three-dimensional  view of
selected magnetic field lines ({\em right}) for the perturbed case.  }
\end{center}
\end{figure}

During the dynamical  evolution of magnetic tower jet,  the narrow ``jet
body'' and  the expanded ``lobe'' are  formed as shown in  Fig.  1 ({\em
left}). The  3D view of  magnetic field lines  as illustrated in  Fig. 1
({\em right}) indicates  that the magnetic tower jet  has a well-ordered
helical field  configuration showing that a tightly  wound central helix
goes up along  the central axis and a loosely wound  helix comes back at
the  outer  edge of  the  magnetic  tower.  The profiles  of  underlying
external gas plays an important role in the transition of jet/lobe.  The
jet body  is confined jointly by  the external pressure  and the gravity
inside $R_{\rm c}$, while  it expands radially beyond $R_{\rm c}$ to
form the lobe in a steeply decreasing ambient pressure.

\begin{figure}[!h]
\begin{center}
\includegraphics[scale=0.32,            angle=0,           clip]{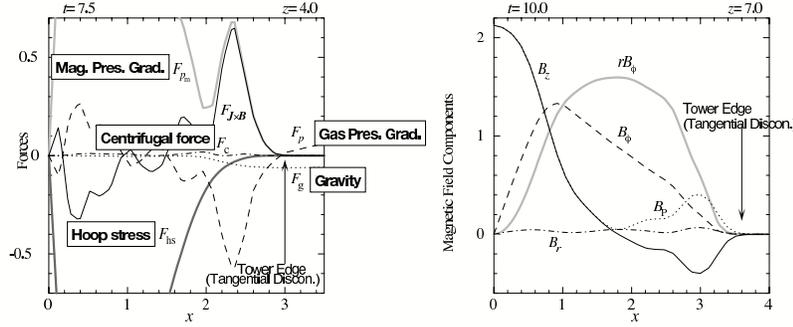}
\caption{\label{fig2} Radial profiles of the forces ({\em left}) and the
magnetic  field components  ({\em  right}) along  the  $x$-axis for  the
unperturbed case. The position of the magnetic tower edge is shown.  }
\end{center}
\end{figure}

The jet  axial current and the  ambient gas pressure  can determine the
radius of  the magnetically dominated lobes \citep[]{N06},  that is, the
internal  gas  pressure  plays a  minor  role  in  the lobes  which  are
typically  seen in  FR I  radio galaxies  \citep[]{C03}.   Our numerical
result shows the magnetically  dominated lobes expand subsonically which
can   be  seen   in  some   observations  \citep[]{K06}.    A  preceding
hydrodynamic shock wave beyond the  tower, which may be identified as an
AGN-driven shock in recent X-ray observations \citep[]{F05, F06}.

\begin{figure}[!h]
\begin{center}
\includegraphics[scale=0.625,          angle=0,            clip]{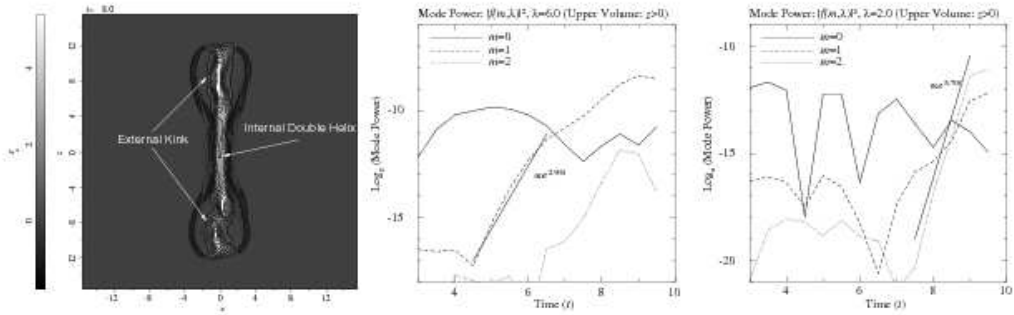}
\caption{\label{fig:f3}  Distribution of  axial current  density  in the
$x-z$  ({\em  left}) for  the  perturbed  case.  Time variation  of  the
azimuthal  Fourier power  of the  current density  with the  axial ($z$)
wavelengths  $\lambda=6.0$   ({\em  middle})  and   $\lambda=2.0$  ({\em
right}).  }
\end{center}
\end{figure}

The  outer edge  of the  magnetic tower may be identified  as a tangential
discontinuity without the normal  field component. The interior of tower
(lobe)  is  separated from  the  non-magnetized  external  gas via  this
discontinuity.  Figure 2 shows  the distributions of forces ({\em left})
and  the  magnetic  field  components  ({\em right})  along  the  radial
directions.  At the tower edge, the outwardly directed magnetic pressure
gradient force  is roughly balanced  with the inwardly  directed thermal
pressure gradient  force. On  the other  hand, at the  core part  of jet
body, the  quasi-force free equilibrium  is achieved. In the  context of
magnetically  controlled  fusion  systems,  the  helical  field  in  the
magnetic  tower  can be  regarded  as  the  reversed field  pinch  (RFP)
profile.

The current-carrying magnetic tower  jet, which possesses a highly wound
helical  field,  is subject  to  the  current-driven instability  (CDI).
Although the destabilizing criteria will  be modified by the ambient gas
and the  RFP configuration  of the tower,  we find that  the propagating
magnetic tower  jets can develop  the non-axisymmetric CDI  modes.  Both
the  internal elliptical  ($m=2$) mode  (Fig.  3  {\em right})  like the
``double  helix'' and  the  external  kink ($m=1$)  mode  (Fig.  3  {\em
middle}) grow to produce the wiggles at the different parts (Fig. 3 {\em
left}) \citep[]{N07}.  This morphological  feature (the jet  is suddenly
disrupted at  the lobe  region with visible  [external] wiggles)  can be
observed in Hercules A \citep[]{GL03}.

\section{Conclusion}
By  performing 3D MHD  simulations, we  have investigated  the nonlinear
dynamics of magnetic tower jets  in the galaxy cluster system. Under the
gravitationally stratified  atmosphere, the  magnetic tower jet  forms a
global current  closure system having  the reverse field  pinch magnetic
configuration. The magnetic  tower jet is well confined  by the external
gas pressure to form the collimated ``jet body'' inside the cluster core
radius.  The  narrow  jet  expands  radially to  form  the  magnetically
dominated ``lobe'' beyond the cluster core radius. Thus, the profiles of
underlying  external gas  play a  important  role in  the transition  of
jets/lobes. 

The current-carrying  tower jet, which possesses a  highly wound helical
magnetic  field, is subject  to the  current-driven instability.  In the
gravitationally stratified atmosphere, the effect of thermal confinement
on the  magnetic tower jet will get  gradually weaker as it  grows. As a
result,  a separation  of current  flowing  path between  the jet  axial
current and the return current occurs. Therefore, the edge of the jet
axial current becomes a free boundary against the external kink mode to
produce the visible wiggles as seen in some observations.

\acknowledgements 
This work  was carried  out under the  auspices of the  National Nuclear
Security Administration of  the U.S. Department of Energy  at Los Alamos
National  Laboratory  under  Contract  No.  DE-AC52-06NA25396.   It  was
supported by the Laboratory Directed Research and Development Program at
LANL and by IGPP at LANL.

\end{document}